# How to Reliably Measure Carrier Mobility in Highly Resistive Lead Halide Perovskites with Photo-Hall Experiment

*Soumen Kundu, Yeswanth Pattipati, Krishnamachari Lakshmi Narasimhan[a], Sushobhan Avasthi\**

Centre for Nano Science and Engineering (CeNSE), Indian Institute of Science, Bangalore 560012 Karnataka, India

[a]Department of Electrical Engineering, IIT Bombay Powai 400076 Mumbai Maharashtra India

AUTHOR INFORMATION

**Corresponding Author**

\* Sushobhan Avasthi- Centre for Nano Science and Engineering(CeNSE), Indian Institute of Science, Bangalore 560012 Karnataka, India https://orcid.org/0000-0001-6201-7711 ; Email: savasthi@iisc.ac.in

**ABSTRACT**

Mobility measurements in highly resistive methylammonium lead iodide (MAPI) are challenging due to high impedance, ion drift, and low mobility. We show that we can address the challenge using intensity-dependent photo-Hall measurements. The key is an improved photo-Hall setup, which enables reliable Hall measurements in the dark and under low-intensity illumination. By tuning the illumination over four orders of magnitude, we get the additional information to simultaneously extract hole mobility, electron mobility, and background doping. For the first time, we show that a MAPI single crystal, exhibiting n-type behaviour in the dark, can show p-type behaviour under light due to the difference in hole and electron mobility. The data partly explains the variability in mobility reported in the literature. We show that one can erroneously extract any mobility from 0 to 25 cm$^2$/Vs if we restrict the experiment to a small range of illumination intensities. For our MAPI (310) crystal, the measured hole and electron mobility is 40 cm²/Vs and 25.5 cm²/Vs, respectively.

MAIN TEXT

Hybrid halide perovskites are remarkable optoelectronic semiconductors with applications extending from thin-film solar cells, X-ray detectors, and light-emitting diodes [1–3]. There is an obvious interest in characterizing the electronic carrier mobility of halide perovskites. Unfortunately, reliable measurements of this fundamental property are challenging for various reasons in halide perovskites. The mobility measured in thin-film transistors underestimates the bulk mobility due to surface scattering and field-effect.[4] Measurements based on space charge limited current (SCLC) characterize only the majority carriers and are affected by shallow traps[5,6]. Hall measurements are difficult in halide perovskites due to high dark resistivity (~ GOhm), large geometric offsets, and low mobility[7]. Pozdorov et al. overcame the limitation of standard Hall experiments by pioneering low-frequency AC Hall measurements for measuring majority carrier mobility[8,9]. They also showed that the Hall experiment under photoexcitation, i.e. photo-Hall, can measure minority carrier mobility.

This paper reports intensity-dependent low-frequency AC photo-Hall measurements in methylammonium lead halide (MAPI) single crystals over four orders of photoexcitation, from 0.03 mW/cm$^2$ to 100 mW/cm$^2$. To extract Hall data at such low photoexcitation, we have implemented several hardware improvements, tackling critical issues like high dark resistance, Faraday induction voltage, noise, etc. The larger range of photoexcitation allows us to observe new physics that can settle some debate on the suitability of Hall measurements in hybrid perovskites. We show that the sign of the Hall voltage in an N-type crystal changes from negative to positive beyond a critical light intensity as if the sample transitions from N-type to P-type doping. We can reliably extract the electron and hole mobilities in MAPI single crystals from the transition, which are 40 cm²/Vs and 25.5 cm²/Vs, respectively. The experiment explains why reported values of Hall mobility in hybrid perovskite are so inconsistent[10]. Depending on the experiment details, like background doping and photoexcitation (intentional or otherwise), one may erroneously extract any Hall mobility from 0 to 25 cm$^2$/vs and claim either P- or N-type doping. Finally, we show that the measurement is repeatable.

In the AC Hall experiment, an alternating magnetic field of frequency ω is applied while a DC longitudinal current (I) flows through the sample. The generated Hall voltage ($V_H$) also oscillates at an angular frequency ω. $V_H$ is measured using a lock-in amplifier with the AC magnetic field as the reference. The measurements were done in a custom-designed setup, similar to that used by Podzorov et al.[8], but with notable improvements. Reliable Hall measurements at low photoexcitation were challenging, requiring several incremental improvements in the measurement circuit. Figure 1b shows the simplified schematic. Details are in Supplementary Section 3. The magnetic field is applied in the vertical direction with permanent N-52 magnets in NS-SN configuration (Figure 1c). The separation between magnets is 27 mm. The magnetic field mechanically oscillates at 1 to 1.5 Hz using a brushless DC (BLDC) motor in closed-loop control. Compared to the alternatives we tested; the BLDC motor provided a more stable frequency with < 1% fluctuation.

A stable frequency is critical to generate a good reference signal for the lock-in amplifier. A linear Hall sensor (AH49H) placed under the magnet provides the reference signal for the lock-in. Although an effort was made to place the Hall sensor precisely at the same position as the sample, the match is imperfect. The field and phase of the magnetic field were calibrated using a standard p-type Si sample with known mobility (see Supplementary Section 4). At 1.5 Hz, the root mean square amplitude of the magnetic field was 0.17 T. The samples were mounted in an inert hermetic chamber, evacuated and filled with $N_2$ to protect the perovskite.

One of the main challenges with AC Hall measurements is induced Faraday voltage due to carrying magnetic flux and noise. The former limits the experiments below < 100 Hz, and the latter causes a parasitic signal above > 1Hz. In our setup, 1 to 1.5 Hz gave the highest SNR. Even at this frequency, a reliable Hall signal could only be obtained with circuit improvements. The noise was controlled by shielding the signal pathways with ground planes. The sample was mounted on a "sample-PCB", whose traces were "twisted" to minimize the inductance and Faraday voltage (Figure 1d).

The resistance of perovskite crystals under low light is ~ 1GΩ, much higher than the input impedance of most lock-in amplifiers. To enable Hall measurements at low photoexcitation, we route the Hall voltage to a differential buffer with an input impedance of $10^3$ GΩ (Figure 1e). The unity-gain buffer is implemented on a second PCB using a CMOS instrumentation amplifier with shielded inputs (INA116UA). The amplifier also converted the differential Hall signal to a single-ended signal with low output impedance. The second PCB had to be > 10 cm away from the magnets to prevent interference.

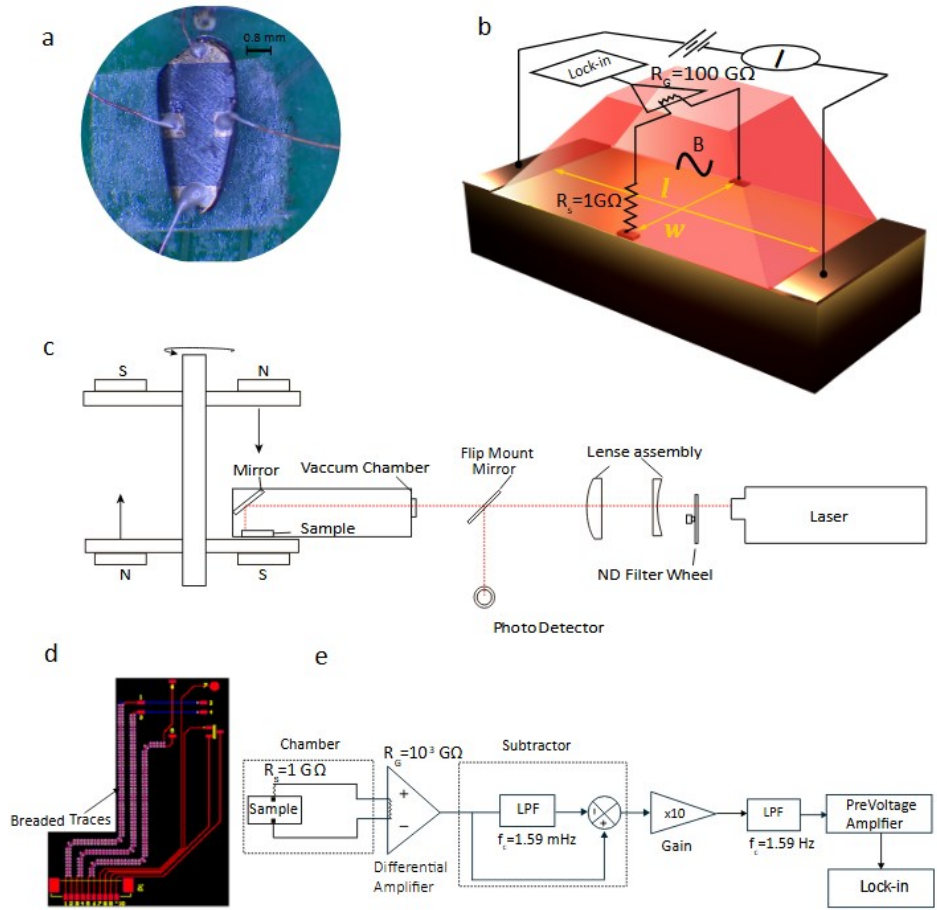

**Figure 1**: Illustration of AC Hall Experiment (a) actual snapshot of Hall device (b) A schematic picture of 4 bar Hall structure under laser illumination with alternating magnetic field (c) A schematic of photo AC Hall setup, laser mounted on XYZ stage followed by ND filter for intensity modulation, laser spot magnifying lens assembly and, on the left vacuum shield chamber in-between rotating magnetic poles (NS-SN type) (d)Sample mounted on a L shaped PCB sticked onto chamber; All Hall experimental leads routed from chamber to external PCB,(e) Overall schematic diagram for Electrical connection, sample mounted on chamber PCB, then routed to offset correction circuit followed by pre-voltage amplifier and lock- in.

The signal from the sample consists of the low-frequency AC Hall signal and a large DC offset due to the misalignment of the Hall electrodes and the thermoelectric effect. The DC offset also drifts with time in perovskites.[11] Both these problems are mitigated using a low pass filter and subtractor, which extracts and cancels the DC component (shown in Figure 1e), yielding the pure AC Hall signal. To further improve SNR, the resultant signal is amplified 10 times and passed through another low pass filter to limit the noise. The amplifier was implemented onboard using OPA2828IDGNT. Finally, a pre-voltage amplifier (SRS-560) was used as a band pass filter with LPF cut-off at 3 Hz and HPF cut-off at 1 Hz with a 6 dB/octave roll-off. The output of the SRS-560 was routed to a lock-in amplifier (SRS865). Typically, 10 s time base and 24 dB/oct roll-off were used in the lock-in.

The photo Hall experiment was conducted using a 634 nm continuous diode laser to obtain both electron and hole mobility. The optical arrangement is shown in Figure 1c. The light passes through a variable neutral density filter lens assembly to vary intensity over four orders of magnitude, from 31 $\mu$W/cm² to 108 mW/cm², i.e. 0.0003 Sun to 1 Sun. The laser intensity was in real-time calibrated using a Si photodetector (SM1PDB) and flip mount mirror. The 4 mm² laser beam was expanded to 16 mm² before entering the vacuum chamber through an optical window, creating a uniform light illumination on the sample.

The crystal was mounted onto a PCB for measurement (Figure 1a). A thin copper wire connects the sample to the PCB pads. Voltage bias ($V_{ap}$) is applied in the longitudinal direction between pad 6 (high) and 5 (ground) using a source measuring unit (Keysight B2902a); simultaneously, it measures current. The applied bias voltage is swept twice to counteract hysteresis, from 1.5 V to 3.0 V and 3.0 V to 1.5 V, with a step of 0.5 V. The Hall signal ($V_H$) is collected from a transverse direction across pads 1 and 2. The reported signal is the average of the two measurements. To counteract drift, we measure the Hall signal only after the Hall current stabilizes at each bias point, which typically takes around 100 seconds (See Supplementary Figure S5). The Faraday signal ($V_F$) due to the time-dependent magnetic field is estimated from the transverse signal at $V_{ap}$=0. The true Hall signal ($V_{Hall}$) is extracted from the measured transverse signal by vectorially subtracting the Faraday component[8].

The θ-2θ X-Ray Diffraction (XRD) pattern of the MAPI crystal shows a single peak at 31.60° (Figure 2(a)). The rocking curve FWHM at 2θ = 31.6° is 0.78°, confirming its single crystalline nature. However, the peak at 31.60° could be due to (111), (114) or (310) planes.[12,13] In order to confirm the orientation of the crystalline surface, Pole Figure (PF) maps for <211> and <110> planes were taken along with the central diffraction spot, whose indices are yet to be confirmed (Figure 2(b)). The obtained PF Maps were compared with WinWulff© simulations (Figure S11) to confirm that the crystal is oriented along (310). Details are in the Supplementary Section 6. Figure 2(c) displays the structural view of MAPI crystal with (310) and (001) planes in blue and pink traces, describing the Hall probe plane and applied bias directions, respectively. The photoluminescence peak at 769nm (Figure 2d), measured with a 400 nm CQ laser, is in good agreement with literature[14,15]

Photoconductance of the crystal was measured by applying a voltage in the longitudinal direction between contacts 5 and 6 in Figure 1b. The photoconductivity $\sigma_{pc} = \sigma_L - \sigma_d$, where $\sigma_L$ and $\sigma_d$ are the areal conductivity under light and dark, respectively. The areal photoconductivity (Figure 2e) has a power law dependence with intensity: $\sigma_{pc} = AI_0^{0.92}$. A near-unity exponent suggests that the photocarrier recombination is monomolecular, like earlier results[16].

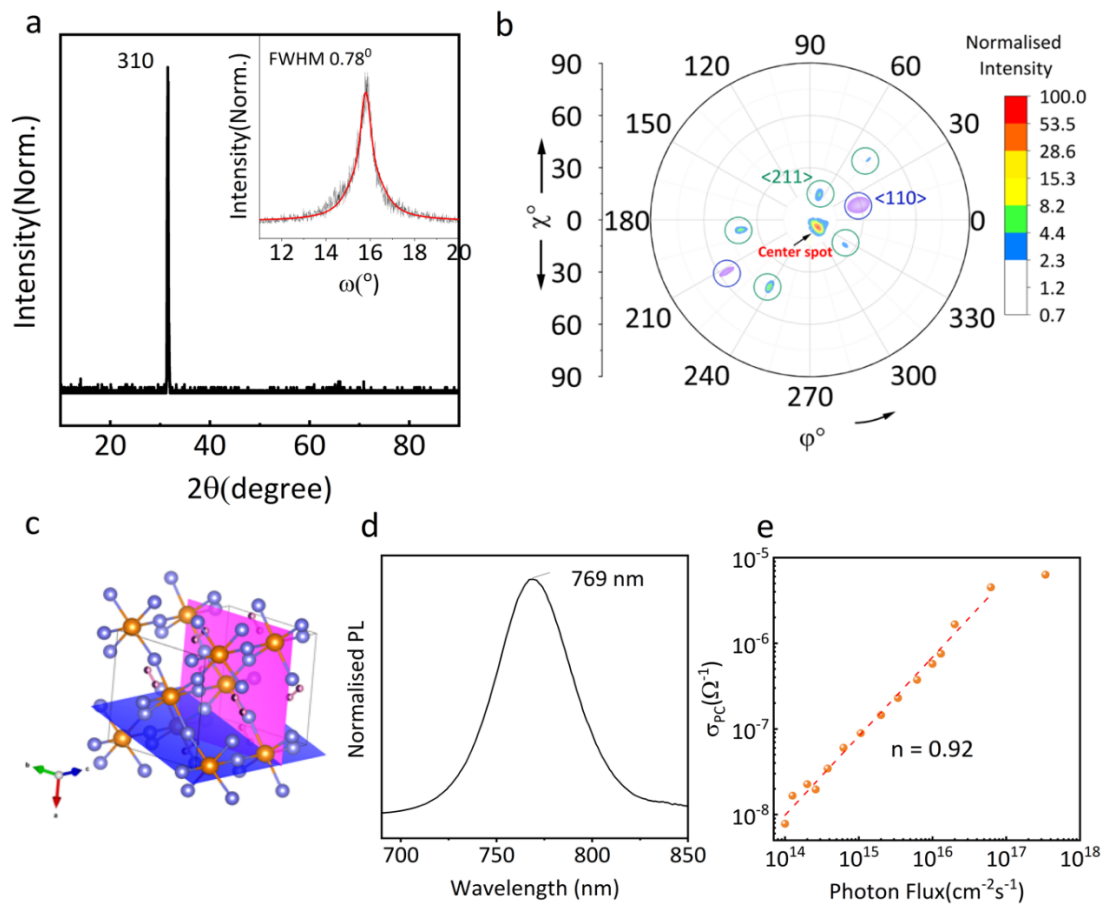

**Figure 2:** Structural and optical characterization of the MAPI single crystal (a) XRD θ-2θ scan of the probing surface used in photo Hall measurement, with an inset of the Rocking curve for the same surface (b) Pole figure Maps of (110), (211) planes of the crystal, along with the central diffraction spot, all overlapped in a single Polar Plot (c) Ball stick model view of MAPI crystal structure with blue plane shows (310) and pink plane shows (001) plane, Pb, I and MA motifs are shown yellow, purple and double bonded brown sphere respectively (d) Photo luminescence emission spectra on the same surface. (e) Sheet photoconductance of MAPI crystal with respect to photon flux

We observed linear current-voltage relation at all intensities (Supplementary section 8), suggesting the contact resistance is negligible.

Beyond $6 \times 10^{16}$ /cm²s the photon flux slope decreases, indicating the onset of bimolecular recombination. Without knowledge of diffusion length, it is difficult to calculate the bulk density of photogenerated carriers. However, we can define areal photoconductance ($\sigma_{pc}$) as:

$$\sigma_{pc} = \frac{1}{wV_{ap}} I_{pc} = e\left(\mu_e + \mu_h\right)\Delta p \qquad (1)$$

Here, $\Delta p = \Delta n$ is the areal excess carrier concentration of photogenerated holes & electrons; $e$ is the electronic charge; and $\mu_h$ & $\mu_e$ are hole and electron mobility, respectively. The distance between longitudinal and transverse contacts are l & w, respectively. We measure dark areal conductance at 2V bias, $\sigma_d = 4.6 \times 10^{-8}$ $\Omega^{-1}$ shown in supplementary section (Figure S14(a)).

Figure 3a shows the measured Hall signal ($V_H$) as a function of longitudinal $V_{ap}$ in the dark and under illumination. The $V_H$ in the dark (shown as a black square) is the lowest curve in the fourth quadrant. The $V_H$ versus $V_{ap}$ slope shows that the background doping in the MAPI crystal is N-type. This also agrees with previous observations in MAPbI$_3$ crystals with 1:1 stoichiometry of MAI & PbI$_2$[17,18]. Hall and conductivity measurements show the electron mobility, $\mu_e = \frac{L \cdot V_H}{B \cdot w \cdot V_{ap}}$, is $23 \pm 0.4$ cm²/Vs and the areal background electron concentration, $n_0 = \frac{\sigma^0}{e\mu_e}$, is $(1.2 \pm 0.4) \times 10^{10}$ cm$^{-2}$. Here, $B$ represents the AC magnetic field's root-mean-square (RMS) amplitude. The phase and magnitude of the magnetic field were independently confirmed by measuring the Hall effect in p-type silicon and n-type Ga$_2$O$_3$ samples (Supplementary Section 6).

Figure 3a also shows the photo-Hall voltage at different light intensities. As the illumination intensifies, the photo-Hall voltage decreases, changing signs when the light intensity is 0.19 mW/cm$^2$. The photo-Hall voltage remains positive at a higher intensity. Figure 3(b) displays photo-Hall resistance ($R_{xy}^{Photo} = \frac{V_{photoHall}}{I}$) as a function of incident photon flux. $R_{xy}^{Photo}$ is negative in dark and low light, goes through zero at a photon flux of 5x10$^{14}$/s.cm$^2$, becomes positive in the interim, and tend towards zero once again at higher light intensity. A trivial analysis of $R_{XY}$ can lead to any value of mobility including zero, depending upon experimental conditions. This probably explains some of the inconsistency people report in Hall measurements in perovskites.

During Photo-Hall experiments, the Lorentz force acts on both electrons and holes, reducing the Hall voltage.[19] Podzorov et al. showed that for an N-type sample, the photo-Hall resistance $R_{xy}^{Photo}$ could be written as:

$$R_{xy}^{Photo} = \frac{B[-n_0\mu_e^2 + \Delta p(\mu_h^2 - \mu_e^2)]}{e[n_0\mu_e + \Delta p(\mu_h + \mu_e)]^2} \tag{2}$$

The change in polarity of $R_{xy}^{Photo}$ observed in Figure 3(b) is a direct consequence of Equation (2). In dark, $\Delta p=0$, Equation (2) simplifies to an n-type semiconductor. Under light, $\Delta p \neq 0$, the $R_{xy}^{Photo}$ changes sign from negative to positive, reaching a maximum some $\Delta p > \Delta p_T$, before falling towards zero at very-high $\Delta p \gg \Delta p_T$.

The generalized photo-Hall mobility ($\mu_{PhotoHall}$) for n-type materials can be extracted from $R_{xy}^{Photo}$ using:

$$\mu_{photo\ Hall} = \frac{l}{B\ w} \frac{V_{PhotoHall}}{V_{ap}} = \frac{-n_0\mu_e^2 + \Delta p(\mu_h^2 - \mu_e^2)}{\Delta p(\mu_e + \mu_h) + \mu_e n_0} \tag{3}$$

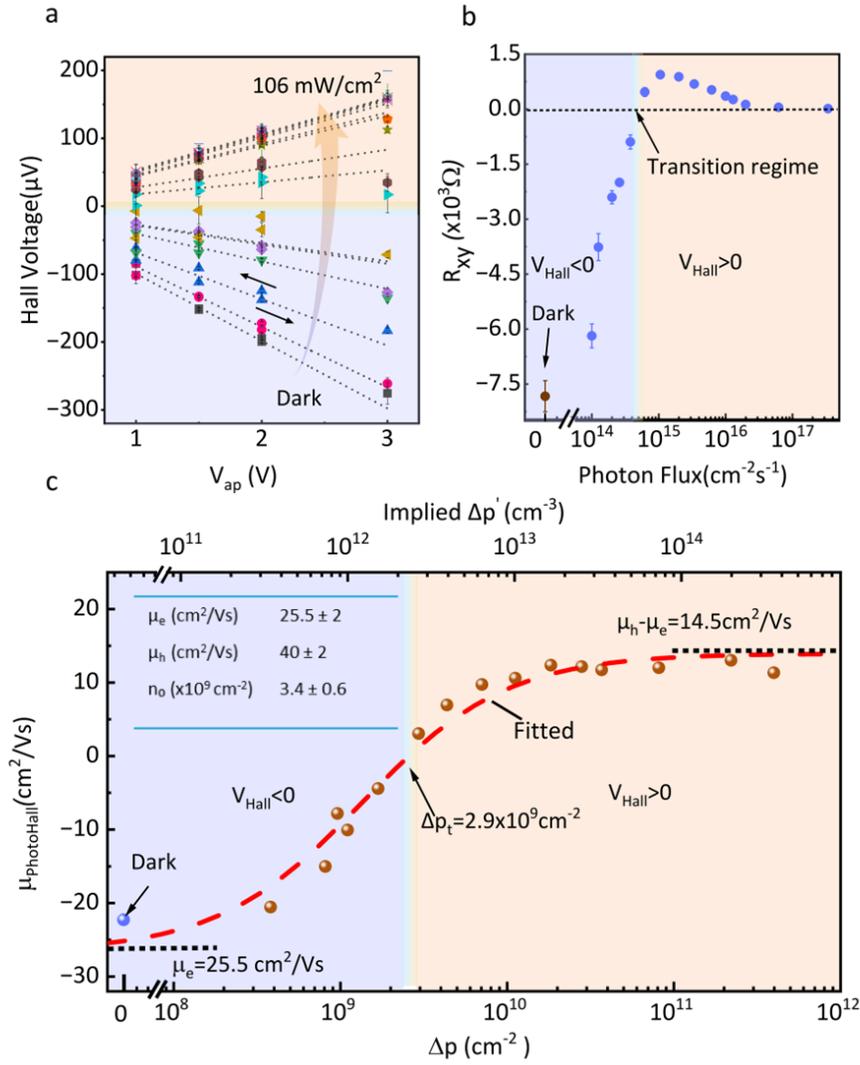

**Figure 3**: The photo-Hall effect of MAPI crystal. a) $V_{PhotoHall}$ signal with respect to $V_{ap}$ is plotted as a function of light intensity varying from 0 mW/cm² to 106 mw/cm². Both forward and backward sweep are given along with standard deviation. b) Hall resistance with respect to photon flux, zero crossing indicates transition from N-type to P-type regime. c) Experimentally measured $\mu_{PhotoHall}$ (brown sphere) as a function excess carrier density (Δp) and implied average carrier concentration. Red dashed line is the fit of equation 3. The extracted values of mobility are also given.

In the dark ($\Delta p = 0$), the $\mu_{photoHall} = -\mu_e$. The negative sign informs that the mobility is for electrons. Under illumination, $\mu_{PhotoHall}$ changes sign at a transition areal density, $\Delta p_t = \frac{n_0}{\left(\frac{\mu_h}{\mu_e}\right)^2 - 1}$. At $\Delta p = \Delta p_T$ the Hall voltage is zero. Under high-level injection, $\Delta p \gg n_0$, $\mu_{photoHall} = \mu_h - \mu_e > 0$. An illustrative simulation of this evolution is given in Figure S2(b).

Simultaneously fitting Equation (1) and Equation (3) with measured data in Figure 3c, we extract three constants: $n_0$, $\mu_e$, & ($\mu_h - \mu_e$). Using numerical iteration, we also extract the relationship between incident photon flux and excess carrier concentration, $\Delta p$ (details in Supplementary Section 9). The evolution of $\mu_{PhotoHall}$ as a function of $\Delta p$ is shown in Figure 3(c). For completeness, the dark Hall data is also shown as a blue marker in the 3rd quadrant. The numerical fit, shown as a red dashed line, is excellent. The extracted $\mu_e$ is $25.5 \pm 2$ cm²/Vs, matching the value extracted in the dark and at low light intensity. At high-level injection ($\Delta p \gg n_0$), extracted $\mu_h - \mu_e$ is $14.5 \pm 2$ cm²/Vs. The resulting $\mu_h$ is $40 \pm 2$ cm²/Vs. The transition in sign occurs at $\Delta p_T = 2.3 \times 10^9$ cm$^{-2}$. The areal background carrier concentration ($n_0$) extracted from $\Delta p_T$, is $(3.4 \pm 0.6) \times 10^9$ cm$^{-2}$, which is close to the value measured in the dark. The areal dark conductivity ($\sigma_0$) calculated from the fitted values of $\mu_h$, $\mu_e$, and $n_0$ is $1.38 \times 10^{-8}$ $\Omega^{-1}$, which is reasonably close to the measured value of $4.6 \times 10^{-8}$ $\Omega^{-1}$ in dark (red square in Figure S14(a)).

Such intensity-dependant photo-Hall measurements over a wide range have distinct advantages over previous measurements[20]. For example, Chen et.al[21] reported photo-Hall measurements above $\Delta p_T$, where $\mu_{photoHall} \approx (\mu_h - \mu_e)$, a constant. We measure photo-Hall at very low intensities, which allows us to observe the complete evolution of photo-Hall mobility from dark to $\Delta p_T$ and above. The physical model in Equation (3) and data in Figure 3(c) provide the information to reliably extract background doping, doping type, electron and hole mobility in the same experiment with more certainty.

The data in Figure 3 also explains the continued debate over the accuracy of Hall experiments in hybrid perovskites. Most experiments are conducted in a dark or small range of light intensities. Depending on the setup and sample, the sign of the signal can also be either negative or positive, and the extracted mobility can vary. For example, with the data in Figure 3, one can extract any photo-Hall mobility from -25 cm$^2$/Vs to 14 cm$^2$/Vs. We avoid the confusion only because we measure the signal over 4 orders of magnitude of photoexcitation, from high-level injection to dark.

Hall measurements at a low light intensity, consistent with low-level injection, are beset by high DC offset, internal resistance, drift, and variability. Solving these issues using the hardware improvement described above is the key to reliable Hall measurements in soft and defective hybrid lead halide perovskites. As mentioned, we extract the relationship between incident photon flux and excess carrier concentration using numerical iteration (Figure S5). The slope is proportional to the $\mu_{eff}\tau^*$ product (details in Supplementary S1), assuming local charge neutrality, strong light absorption, and long diffusion lengths. Here, $\tau^*$ is the effective lifetime incorporating bulk and surface effects. The mobility, $\mu_{eff} = \mu_e + \mu_h$. From Figure S5, we extract a $\mu_{eff}\tau^*$ of $(2.2 \pm 0.04) \times 10^{-4}$ cm$^2$/V. Given the mobility, the $\tau^*$ is $3.4 \pm 0.27$ µs, matching previously published data[22]. The carriers diffuse vertically from the front surface into the bulk. This diffusion is probably ambipolar with a diffusion coefficient of $D=\frac{D_n D_p}{D_n+D_p}$. Assuming Einstein's relationship between D and $\mu$, we extract an ambipolar diffusion length of $L_{diff} = 11$ µm. The implied average bulk carrier concentration ($\Delta p' = \Delta p/L_{diff}$) is given in the second X-axis of Fig 3c.

Perovskite crystals are susceptible to drift and degradation, some of which are reversible[23]. Debye layers are formed at the contacts due to halide ion migration within minutes and cation migration over longer timescales.[24,25] Prolonged illumination and applied bias needed for the photo-Hall experiments accelerate these issues. The variability in data during and between experiments is a significant challenge for Hall experiments.

Figure 4 shows the photoconductivity and photo-Hall data for the same crystal over 3 months, during which we refined out methods. We probed the MAPI single crystal for several hours in each run. Multiple runs were conducted over 2 months to improve the methodology and reduce errors. Between these runs, the sample was stored in an $N_2$-filled glove box. The crystal surface was polished with sandpaper, and fresh metal contacts were deposited before each run to mitigate surface degradation.

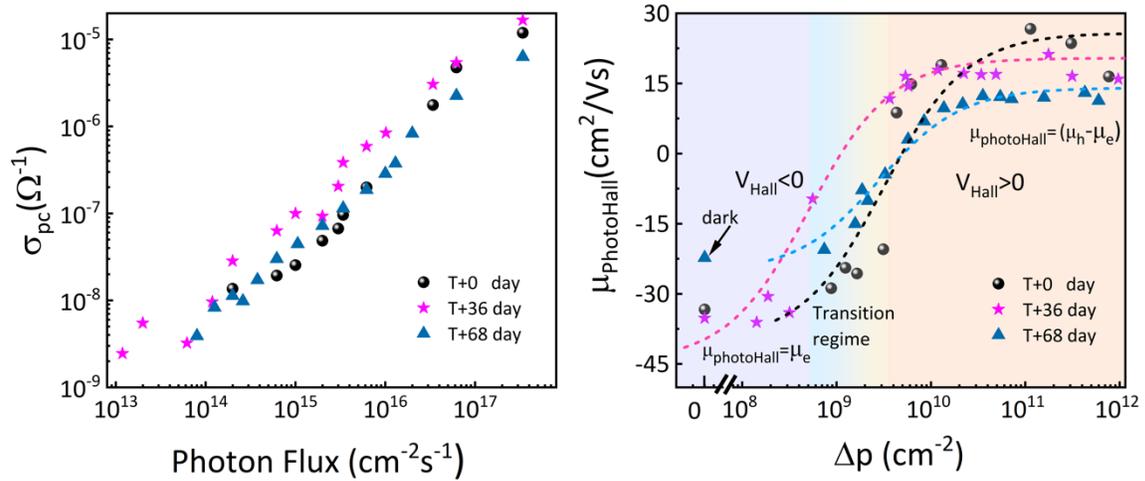

**Figure 4**: MAPI crystal degradation and repeatability study using Photo Hall (a)All three runs' photo conductance with respect to photo. (b) All three runs' experimental $\mu_{PhotoHall}$ and their theoretical fits are plotted with respect to $\Delta p$

A representative summary of three such runs is presented in Table 2. The $\mu_e$, $\mu_e$, $\mu_h$, & $\mu_{eff}\tau^*$ are consistent between Run 1 and Run 2. The repeatability is excellent, especially given that MAPI is unstable. The repeatability shows the robustness of our intensity-dependant photo-Hall approach. Run 3 provided the best quality of data, which we presented in the section above. However, despite our care, the repeated mounting and exposure to ambient over 68 days, resulted in a lower $\mu\tau^*$ (Figure 4(a)) and carrier mobility (Figure 4(b)). Such small changes in photo-physics are usually hard to measure. The intensity-dependent photo-Hall technique is a rich method to monitor degradation in lead halide perovskites.

*Table 2:* Summary of extracted parameters from MAPI single-crystal

|  | **Run 1: T + 0 days** | **Run 2: T+ 36 days** | **Run 3: T + 68 days** |
| --- | --- | --- | --- |
| $\mu_h$(cm$^2$/vs) | 66 ± 8 | 64 ± 8 | 40 ± 2 |
| $\mu_e$(cm$^2$/vs) | 40 ± 9 | 41 ± 9 | 25.5 ± 2 |
| $n_o$ (x10$^9$ cm$^{-2}$) | 8.4±4 | 1.2±0.6 | 3.4±0.6 |
| $\tau^*$ (μs) | 4.1±0.87 | 5.2±0.87 | 3.4 ± 0.27 |
| $\mu\tau^*$ (x10$^{-4}$ cm$^2$/V) | 4.4 ± 0.23 | 5.5 ± 0.03 | 2.2 ± 0.04 |

In conclusion, we report Hall measurements in an n-type MAPI single crystal in dark and under illumination, at intensities from 0.03 mW/cm$^2$ to 100 mW/cm$^2$. We overcame the challenges of Hall measurement in highly resistive perovskite samples by improving the hardware used for the measurements. We used a closed-loop controlled BLDC motor to reduce jitter, implemented an onboard differential buffer to increase the input impedance, minimised the inductance of the circuit to reduce Faraday voltage, and cancelled the dynamic geometric offsets. We find that the Hall voltage signal changes with light intensity -- in our sample from N-type in the dark to P-type above 0.19 mW/cm$^2$. The sign change is because the Hall resistance is a weighted average of electron and hole mobility, which are different in MAPI. Combined with photoconductivity, the evolution of photo-Hall resistance with light intensity gives a direct measure of electron mobility, hole mobility, and background carrier concentration. For the n-type MAPI (310) single crystal, the hole mobility is $66 \pm 8$ cm$^2$/Vs and electron mobility is $40 \pm 9$ cm$^2$/Vs along (001). The data is repeatable for up to a month of repeated measurements. After 68 days of repeated measurements, the hole and electron mobility reduced to 40 cm²/Vs and 25.5 cm²/Vs, respectively. The final background areal carrier concentration is 3 x 10$^9$ cm$^{-2}$. Given the measured values of mobility and background concentration, the predicted dark conductivity matches independent measurements. The implied ambipolar diffusion coefficient is 0.4 cm$^2$/s, which matches previous results. With additional assumptions, we find the diffusion length and carrier lifetime in our sample to be 11 μm and 3.4 μs, respectively.

# Experimental Details

The MAPI single crystal was grown using an inverse temperature crystallization[5]. To prepare 1 M solution, 795 mg methylammonium Iodide (MAI) and 2305 mg of lead iodide (PbI$_2$) were dissolved in 5 ml γ-butyrolactone (GBL) and stirred overnight at 60$^0$C. The solution was filtered with a 0.22 μm PTFE filter and sealed in a vial with parafilm. The vial was gradually heated at $5^0$C/hour in an oil bath. Nuclei of MAPI started precipitating out of the solution at 110 $^0$C. From here on, the temperature was kept constant. Over time, the nuclei grew, forming single crystals of a few mm in size. Crystals with a pseudo-rectangular shape were selected and buffed with sandpaper of different grits ranging from coarse to fine to form a flat facet. Gold contacts were deposited by evaporating through a shadow mask to form Ohmic contacts on the crystal. The longitudinal length between the contacts ($l$) is 4 mm, and the transverse width ($w$) is 1 mm. Figure 1a shows a picture of the sample, and Figure 1b shows a schematic with contacts.

The experimental conditions for XRD and Photo luminescence measurements are discussed in the supplementary section.

ASSOCIATED CONTENT

Supporting information from the authors is available online.

AUTHOR INFORMATION

Soumen Kundu - Centre for Nano Science and Engineering (CeNSE), Indian Institute of Science, Bangalore 560012 Karnataka, India https://orcid.org/0000-0002-4992-1148; Email: soumenkundu@iisc.ac.in

Yeswanth Pattipati- Centre for Nano Science and Engineering (CeNSE), Indian Institute of Science, Bangalore 560012 Karnataka, India https://orcid.org/0009-0003-9151-7805; Email: yeswanthp@iisc.ac.in

Krishnamachari Lakshmi Narasimhan – Department of Electrical Engineering, IIT Bombay Powai 400076 Mumbai  Maharashtra India; Email: kln1948@gmail.com

Sushobhan Avasthi- Centre for Nano Science and Engineering(CeNSE), Indian Institute of Science, Bangalore 560012 Karnataka, India https://orcid.org/0000-0001-6201-7711 ; Email: savasthi@iisc.ac.in

AUTHORS' CONTRIBUTION

SK: Conceptualization, Design of Experiment, Methodology, Investigation, Formal Analysis, Data curation, Visualization, writing original draft

YP: Data curation, Formal analysis

KLN: writing draft, Formal Analysis, Supervision, Conceptualization, Validation

SA: Resources, Conceptualization, Supervision, Project administration, Validation, writing and reviewing.

**Notes**

The authors declare no conflict of interest.

ACKNOWLEDGMENT

Authors are indebted to the Department of Science and Technology (DST), Government of India, under grant DST/ETC/CASE/RES/2023/02. The work was also partially supported by the Ministry of Education, Government of India grant MoE-STARS/STARS-1/135. The work was conducted at the facilities of CeNSE and the Indian Institute of Science (IISc) with the support of the Ministry of Education (MOE) and Ministry of Electronics and Information Technology (Meity), Department of Science and Technology (DST), We thank Dr. Vladimir Bruevich from Department of Physics, Rutgers University for guidance on the experimental setup.

Supporting Information for

# How to Reliably Measure Carrier Mobility in Highly Resistive Lead Halide Perovskites with Photo- Hall Experiment

*Soumen Kundu[1], Yeshwanth Pattipati, Krishnamachari Lakshmi Narasimhan, Sushobhan Avasthi[*]*

Centre for Nano Science and Engineering (CeNSE), Indian Institute of Science, Bangalore 560012 Karnataka, India

[α]Department of Electrical Engineering, IIT Bombay Powai 400076 Mumbai  Maharashtra India

AUTHOR INFORMATION

**Corresponding Author**

* Sushobhan Avasthi- Centre for Nano Science and Engineering(CeNSE), Indian Institute of Science, Bangalore 560012 Karnataka, India https://orcid.org/0000-0001-6201-7711 ; Email: savasthi@iisc.ac.in

# Theory: Photocurrent ($I_{PC}$) in a thick single crystal

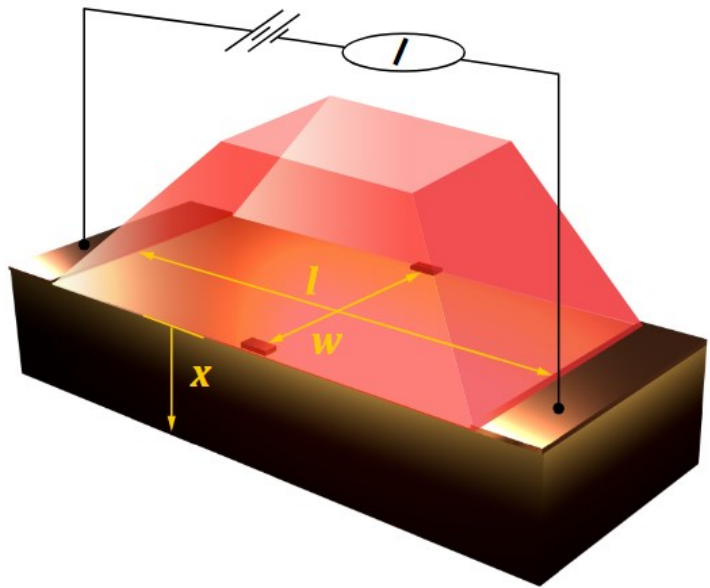

**Figure S1:** *A schematic diagram of a semiconducting crystal under continuous illumination. In the depth (x) how carrier dynamics is behaving has been derived from continuity equation under steady state condition.*

The electric field is assumed to be negligible between the metal contacts shown in Figure S1. At steady-state conditions under illumination, the continuity equation is

$$G(x) - \frac{\Delta p(x)}{\tau} + D \frac{d^2 \Delta p(x)}{dx^2} = 0 \quad \text{(S1)}$$

G(x) is the generation rate, $G(x) = g_0 \cdot e^{-\alpha x}$ where $g_0$ is the generation rate just near the surface. $R(x) = \frac{\Delta p(x)}{\tau}$ is the recombination rate at x from the top surface of the crystal. Excess carrier concentration in ($\Delta p$) in volume diffuses from the top illuminating surface to the depth (x) of the crystal with carrier diffusion coefficient D. Assuming uniform illumination, and $\tau$ is bulk carrier lifetime. If s be the surface recombination velocity for the top and bottom surface and thickness is infinite compared to absorption depth ($\frac{1}{\alpha}$), the solution for the above equation with boundary condition is [1]

$D \cdot \frac{d\Delta p}{dx} = s\Delta p$ at x =0, and $\Delta p = 0$ at x=∞

$$\Delta p = \frac{g_0 \cdot L^2}{D} (1 - \alpha^2 L^2)^{-1} \left[ e^{-\alpha x} - \left( \frac{\alpha + \frac{s}{D}}{L^{-1} + \frac{s}{D}} \right) e^{-x/L} \right] \qquad (S2)$$

Where $L = \sqrt{D\tau}$ is the diffusion length.

Under illumination $I_{pc} = I - I_{dark}$ and under bias voltage V applied on crystal with width w, thickness t and length $l$, $\Delta\sigma$ is bulk pure photoconductivity

$$I_{pc} = \frac{V}{l} tw\Delta\sigma = \frac{V}{l} twe\mu\Delta p_{avg} \qquad (S3)$$

Where $\Delta p_{avg} = \frac{1}{t}\int_0^\infty \Delta p \, dx$ and $\mu$ is effective carrier mobility.

$$I_{PC} = A\frac{V}{l} tweg_0\mu \frac{1}{t}\int_0^\infty \Delta p \, dx \qquad (S4)$$

Assumed $(\alpha L)^2 \gg 1$, $1/\alpha \ll \frac{(S+\alpha D)L}{S+D/L}$ and $g_o = I_0 \alpha$, where $I_0$ is the photon flux on the top surface of the crystal; we consider carrier generation to be 100 % from each photon.

$$I_{pc} = \frac{V}{l} w e I_0 \mu \tau^* = \sigma_{pc} \frac{V}{l} w \quad \text{since, } \sigma_{pc} = t_{ef} \Delta\sigma \text{ and } \tau^* = \tau \frac{S/\alpha + D}{SL+D} \quad (S5)$$

where $\tau^*$ is an effective lifetime which encompasses the effect of surface, and $\sigma_{pc}$ is areal photoconductance.

Effective thickness or ambipolar diffusion length $L_{diff} = \sqrt{D\tau^*}$ gives an idea about how much thickness a crystal block conducts under illumination. Here, $D = \frac{D_n D_p}{D_n + D_p}$, $D_n, D_p$ are electron and hole diffusion coefficient. We extract the ambipolar diffusion coefficient using Einstein's relationship $D = \frac{kT}{e}\mu$.

## Theory: Photo Hall in N-type semiconductor

In a model for the photo-Hall signal in n-type semiconductors proposed by Podzorov et al., the photo-Hall carrier density ($n_{photoHall}$) is determined by the experimentally measured Hall signal ($V_{photoHall}$), the input longitudinal current (I), and the applied magnetic field (B). This photo-Hall carrier density is further related to the excess carrier concentration ($\Delta p$), the dark carrier concentration ($n_o$), and the electron ($\mu_e$) and hole ($\mu_h$) mobilities. Under the assumption that $\Delta p = \Delta n$, the specific relationships between these parameters are described by the following equations

$$n_{photoHall} = \frac{IB}{e \cdot V_{PhotoHall}} = \frac{[n_0 \mu_e + \Delta p(\mu_h + \mu_e)]^2}{[-n_0 \mu_e^2 + \Delta p(\mu_h^2 - \mu_e^2)]} \quad (S6)$$

Equation S6 can be used to derive the generalized Hall resistance under illumination, $R_{xy}^{photo}$, as follows:

$$R_{xy}^{photo} = \frac{V_{PhotoHall}}{I} = \frac{B[-n_0\mu_e^2 + \Delta p(\mu_h^2 - \mu_e^2)]}{e[n_0\mu_e + \Delta p(\mu_h + \mu_e)]^2} \tag{S7}$$

where: e is the charge of an electron, $\mu_e$ and $\mu_h$ are the electron and hole mobilities, respectively.

This model makes the following assumptions: Both electron and hole mobilities are assumed to be independent of the excess carrier concentration ($\Delta p$).

Podzorov et al.[2] described the mobility under illumination ($\mu_{PhotoHall}$) as follows

$$\mu_{PhotoHall} = \frac{L \cdot V_{PhotoHall}}{B \cdot w \cdot V_{ap}} = \frac{-n_0}{\Delta p} \frac{\mu_e^2 + \Delta p(\mu_h^2 - \mu_e^2)}{(\mu_e + \mu_h) + \mu_e n_0} \tag{S8}$$

'w' and 'l' represent the distances between electrodes along the width and length of the semiconductor crystal, respectively (Figure 1). A longitudinal voltage ($V_{ap}$) is applied across the length (l), while the Hall signal is measured across the width (w). Two-probe measurements are considered here, assuming negligible contact resistance compared to the bulk resistance.

The total conductivity ($\sigma$) comprises two components: dark conductivity ($\sigma_o$) and photoconductivity ($\sigma_{pc}$) due to illumination, given by $\sigma_{pc} = e\Delta p (\mu_e + \mu_h)$. The total conductivity is also related to the photo-Hall carrier density ($n_{photoHall}$) and the photo-Hall mobility ($\mu_{photoHall}$) as follows: $\sigma = \sigma_o + \sigma_{pc} = n_{photoHall} * \mu_{photoHall}$.

Equations S7 and S8 are simulated using the following electron and hole mobility pairs: $\mu_h = 10$, $\mu_e = 6.5$ and $\mu_h = 6.5$, $\mu_e = 10$, with a dark n-type carrier concentration of $10^{10}$ cm$^{-2}$.

Analysis of Equation S8 reveals the following trends: at very low light intensities ($\Delta p \ll n_o$), the magnitude of $\mu_{PhotoHall}$ approaches $\mu_e$. As the light intensity increases ($\Delta p \gg n_o$), the value of $\mu_{PhotoHall}$ saturates to ($\mu_h - \mu_e$).

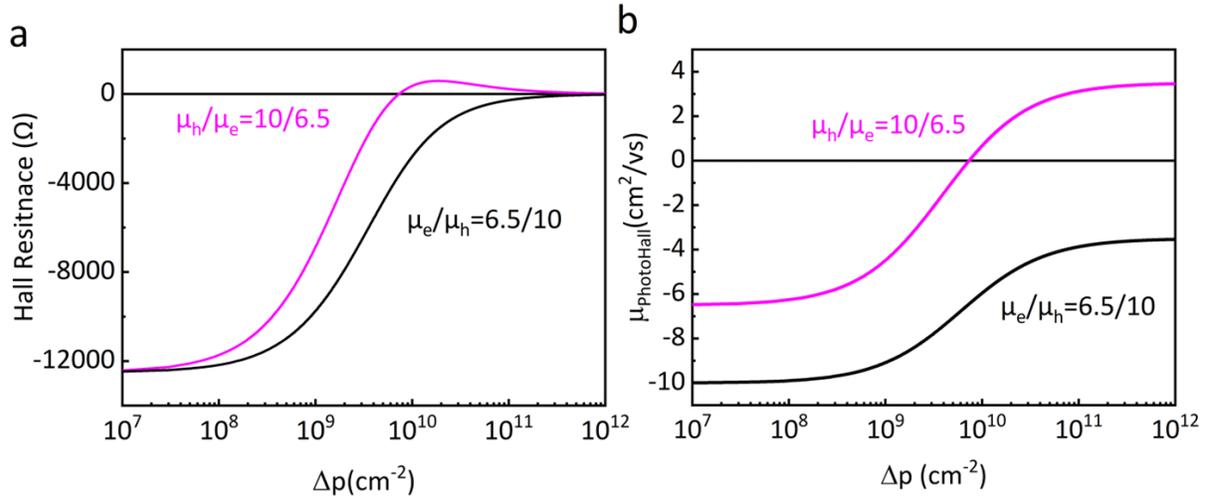

**Figure S2**: *Photo Hall effect of n type semiconductor. In one case electron and hole mobilities are assumed 10 cm²/vs, 6.5 cm²/vs respectively and other case electron and hole mobilities are assumed 6.5 cm²/vs, 10cm²/vs respectively, in both cases $n_0$ is considered $10^{10}$cm⁻² a) Simulation of photo Hall resistance with excess carrier concentration($\Delta p$) given by equation (S7) with two cases discussed above are shown. b) Simulation of $\mu_{PhotoHall}$ as a function of $\Delta p$ given equation S8*

In an n-type semiconductor under dark conditions, if hole mobility ($\mu_h$) exceeds electron mobility ($\mu_h > \mu_e$), the photo-Hall mobility ($\mu_{PhotoHall}$) and Hall resistance ($R_{xy}$) exhibit a sign reversal from negative to positive, as depicted by the pink curves in Figure S2(a) and S2(b). Conversely, if hole mobility is lower than electron mobility ($\mu_h < \mu_e$), the Hall resistance does not cross zero and retains its sign, as shown by the black curves in Figure S2(a) and S2(b). This sign change corresponds to a 180-degree phase shift in the photo-Hall voltage ($V_{PhotoHall}$).

The $\mu_{PhotoHall}$ approaches zero when the photo-Hall signal ($V_{PhotoHall}$) becomes zero at a specific transition excess carrier concentration (($\Delta p_t = \frac{n_0}{1-\frac{\mu_e}{\mu_h}}$)). This occurs when the numerator of Equation S8 becomes zero. Dark Hall measurements provide the majority carrier mobility ($\mu_e$), while in the excitation limit ($\Delta p \gg n_o$), $\mu_{PhotoHall}$ converges to ($\mu_h - \mu_e$).

## Hardware: Subtracting the dynamic drift

The high source resistance and ion migration in perovskites drift the measured signal at various timescales. A dynamic feedback loop-controlled circuit was employed to mitigate this drift in the Hall signal. This circuit was designed around an instrument amplifier with a low input impedance current, as illustrated in Figure S4.

The instrument amplifier ensured that the device's source resistance ($R_S$) under test was at least 100 times lower than the input impedance ($R_G$) of the amplifier, as depicted in Figure S3. Circuit simulation results are briefly discussed in Figure S4. A slow-varying differential input test signal was applied to evaluate the circuit's performance. This test signal consisted of a large offset voltage (0.5 V) superimposed with a small amplitude sinusoidal signal (0.005 V). The sinusoidal signal was designed to exhibit a linear drift over time.

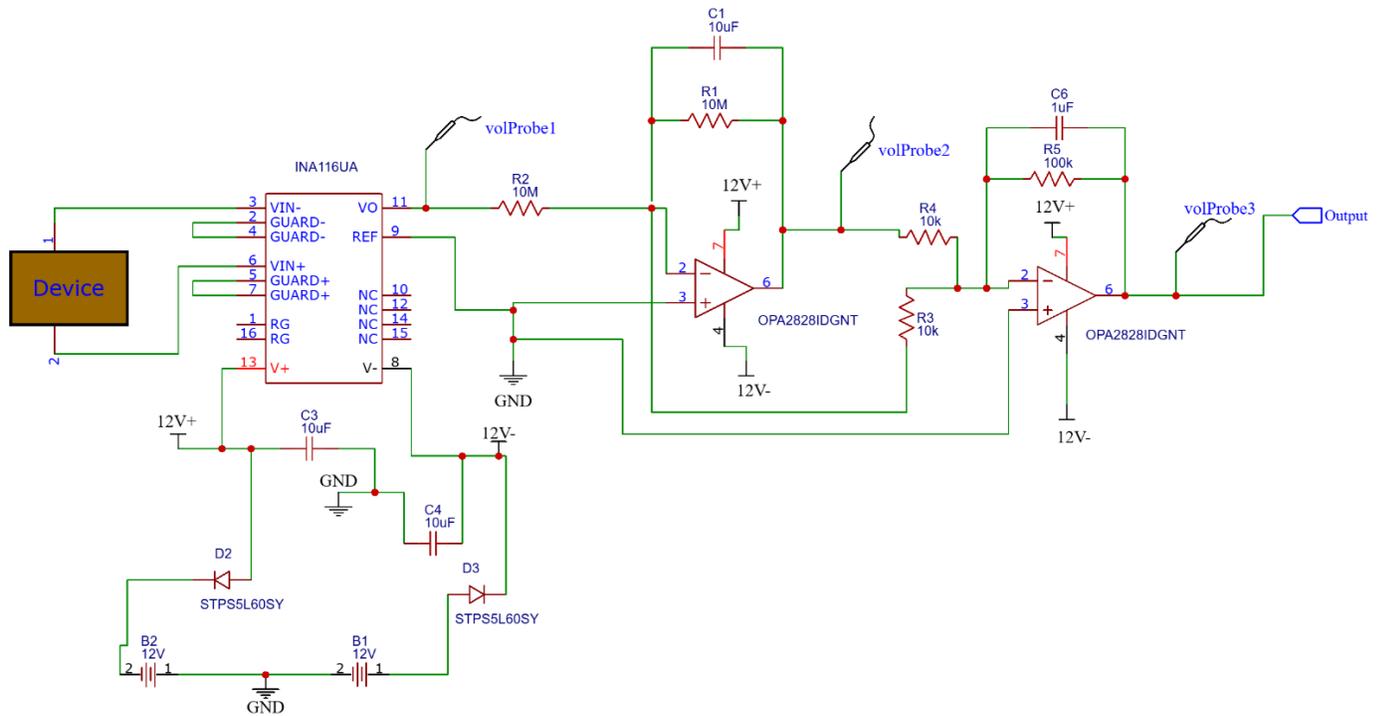

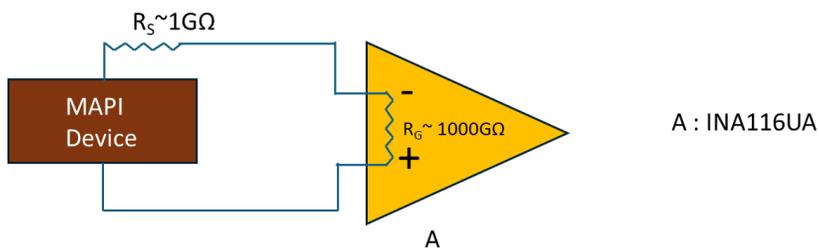

**Figure S3:** *Detailed circuit diagram for dynamic drift cancelation and solution to tackle high source impedance device*

In the inset of Figure S4, magnified views of the input and output signals between 100s and 150s are presented. Despite the input signal's time-varying nature, the circuit's robust design effectively nullifies the drifting offset through the feedback loop mechanism. It is important to note that drift is not solely observed in the Hall voltage but manifests in the Hall current.

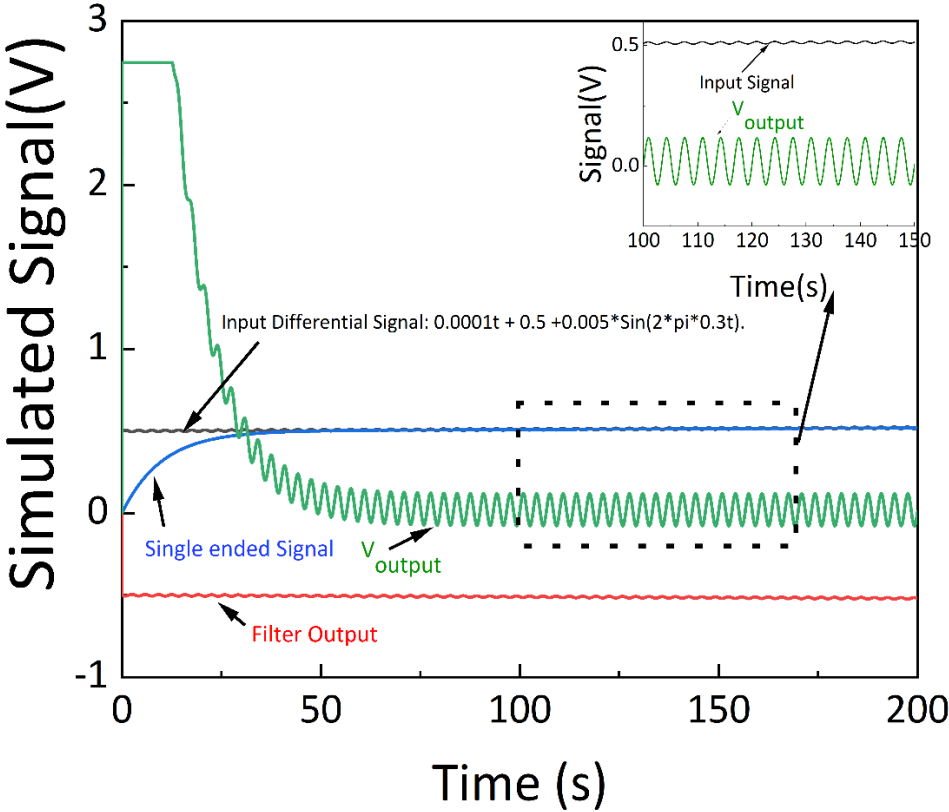

**Figure S4:** *Simulation of dynamic drift cancelation circuit at different nodes with a input test signal like Y= 0.0001t + 0.5 +0.005\*Sin(2\*pi\*0.3t).*

Figure S5 illustrates the current's transient response as a time function. Our observations indicate that the Hall current requires at least 100 seconds to reach a steady-state after applying the bias.

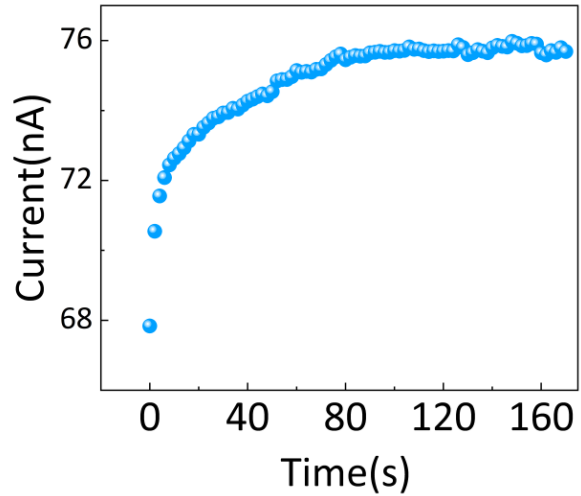

**Figure S5:** *Hall current as a function of time at 3V bias on perovskite sample under $3.7 \times 10^{14}$ $cm^{-2}s^{-1}$ photon flux*

# Hardware: Calibrating polarity by p-type Si and n-type $Ga_2O_3$

Calibration measurements were used to determine the sign of the AC Hall signal in our experimental setup, using standard p-type Si and n-type $Ga_2O_3$. These measurements were conducted with the existing electronic circuitry employed for photo-AC Hall measurements on MAPI perovskite crystals.

Figure S6 presents the AC Hall signal as a function of the applied bias for the p-Si and n-$Ga_2O_3$ samples. As expected, the p-type Si sample exhibited a positive slope in the Hall signal with respect to the applied bias. In contrast, the n-type $Ga_2O_3$ sample displayed a negative slope.

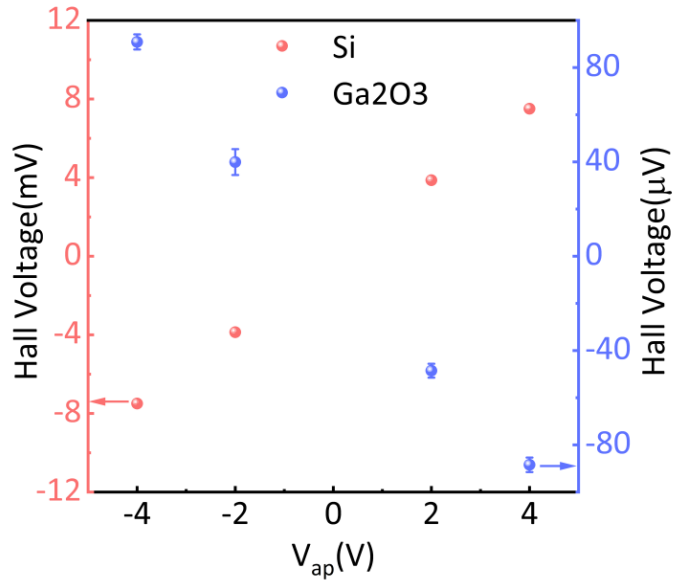

**Figure S6:** *AC Hall signal without illumination in same setup with the existing electronic circuitry. AC Hall voltage as a function of applied voltage pale pink used for p- Si and blue sphere for n-$Ga_2O_3$*

# Hardware: Calibrating magnetic field in the AC Hall setup using p-type Si

The mobility of a standard P-type Si sample can be independently measured from DC Hall measurements. The value measured can then be used to calibrate the magnetic field of our AC Hall setup. We calibrated the magnetic field of the AC Hall setup using a known standard p-type Si sample (resistivity: 1-10 Ω-cm) with a thickness of 330 microns. The sample was diced into a rectangular shape measuring 6 mm by 1.5 mm². The distance between pads 2 and 4 was 4 mm,

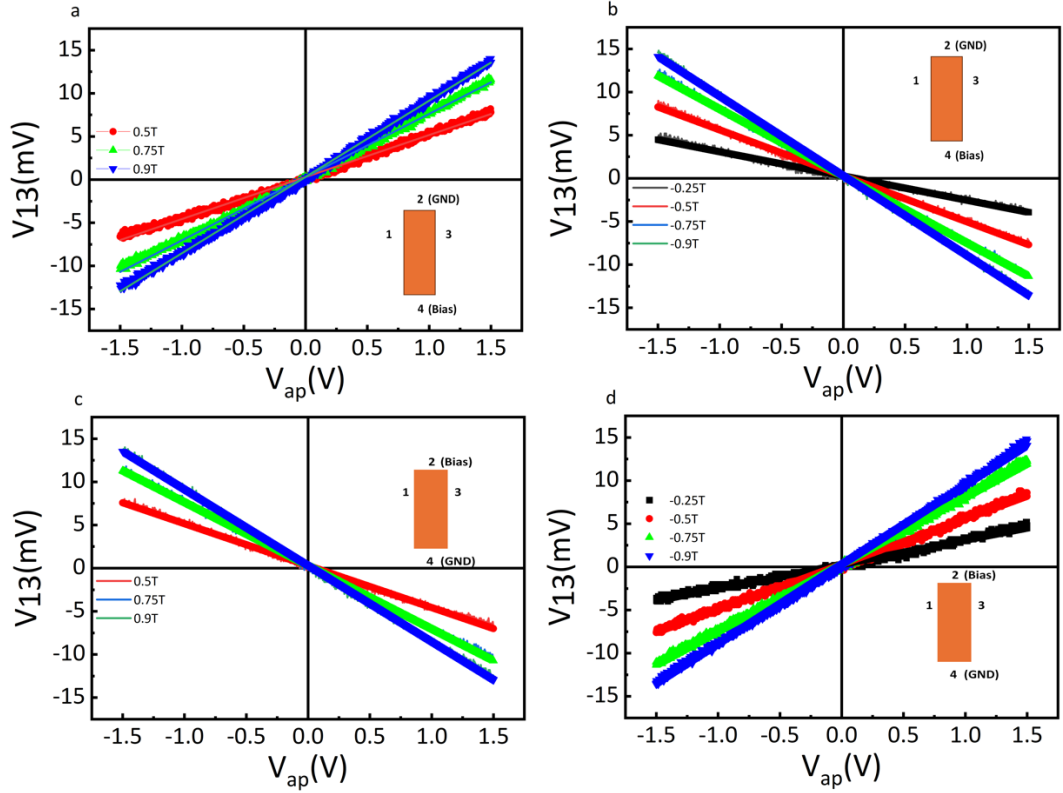

**Figure S7:** *Measured Hall voltage ($V_1$- $V_3$) as a function of applied bias voltage with magnetic fields -0.25T,±0.5T,±0.75T and ±0.9T (a) and (b) figures show applied bias at 4 and ground is at 2 (c) and (d) figures show applied bias at 2 and ground is at 4*

and the width between pads 1 and 3 was 1.2 mm. A 375 nm thick aluminium layer was deposited on the top surface of the sample, followed by rapid thermal annealing at 600 °C. The ohmic nature of the contacts was confirmed by verifying the linearity of the current-voltage sweep. The mobility of the sample was estimated using the following standard expression for Hall bar geometry: [3]

$$V_m = V_{ap}\frac{w}{l}\mu B + V_{ap}\frac{w\alpha}{l} + V_{TH}$$

where $V_m$ is the measured voltage across contacts 1 and 3, $V_{ap}$ is the applied bias across contacts 2 and 4, 'μ' is the mobility, 'B' is the magnetic field, ' w ' is the width, ' l ' is the length, 'α' is the geometric offset, and $V_{TH}$ is the thermal offset between the two Hall contacts.

By analysing the slopes of the data in Figure S7 at different magnetic fields, we can extract (W/l) *(μB + α), eliminating the $V_{TH}$ component. The extracted data was then plotted as a function of the magnetic field (B) in Figure S8.

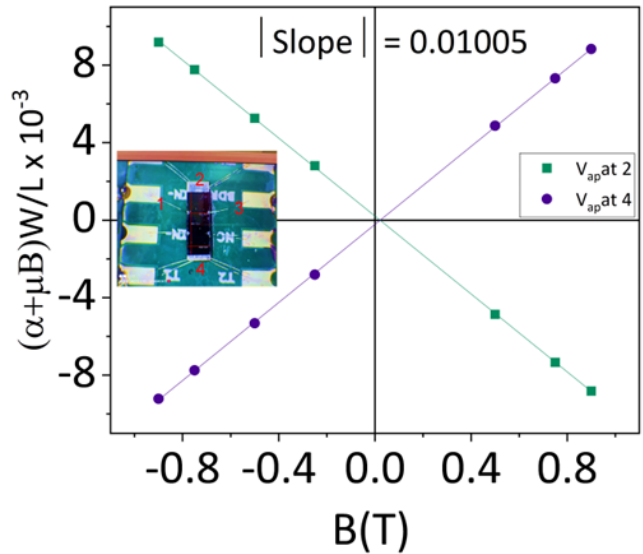

**Figure S8:** *Extracted slope $\frac{w}{l}(\mu B+\alpha)$ is plotted as a function of magnetic fields for two cases one, applied bias on 2 and ground 4 and other applied bias on 4 and ground is 2*

The polarity of the applied bias was alternated to account for any asymmetry between contacts 2 and 4. The average mobility value obtained from this procedure was 335 cm²/Vs. This experimentally determined mobility value for the p-type Si sample was then used to calculate the magnetic field in the AC Hall setup. From the slope ($w / l * \mu B$) of 0.00174 obtained from the AC Hall measurement of the Si sample in Figure S8, and by substituting the known mobility value, the magnetic field was determined to be 0.17 T.

The Si standard can also be used to estimate the repeatability of the setup. Two sets of data measured weeks apart are shown in Figure S9 for the same P-Si standard at 1.5 Hz frequency. The obtained slopes (W/L μB) from Figure S9 give mean $B_{rms}$ = 0.17 T.

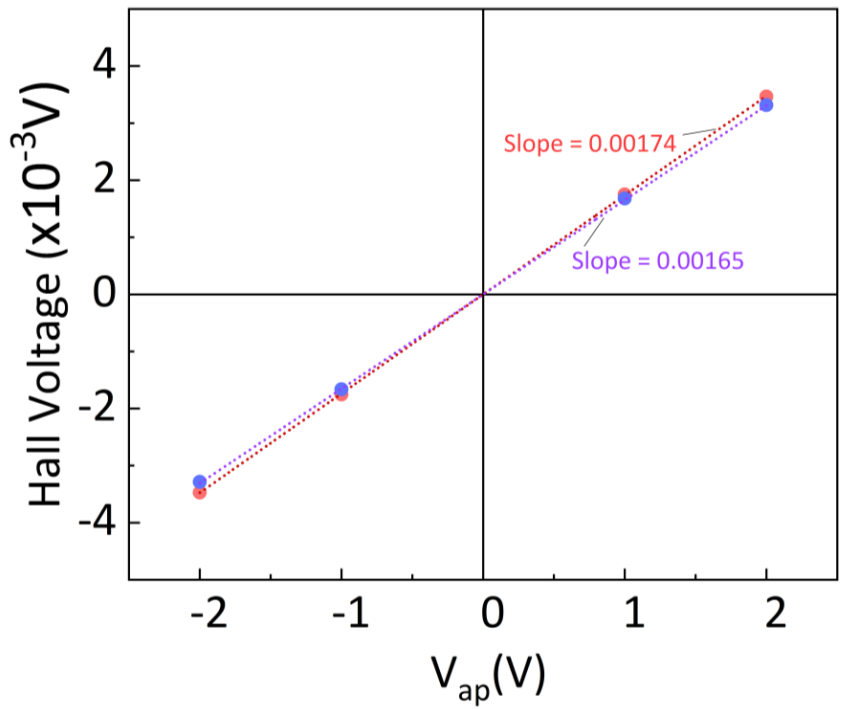

**Figure S9:** *AC Hall voltage as a function of applied voltage for p- Si in two instances*

# Data: Pole-Figure of MAPI single crystal

XRD was done in a Rigaku Smart Lab diffractometer equipped with a vacuum-sealed Cu X-ray source operating at 1.2 kW (40 kV, 30 mA). A 5 mm slit was used on the beam incident side. A single pixel detector was used. A parallel slit analyser was kept open on the detector side. A θ-2θ scan was performed from 10° to 90° at a scan speed of 5°/min with a step size of 0.02°. An omega scan was acquired for the rocking curve while keeping 2θ fixed at the dominant peak. The source and sample holding stage were synchronously rocked along an arc with the φ position fixed.

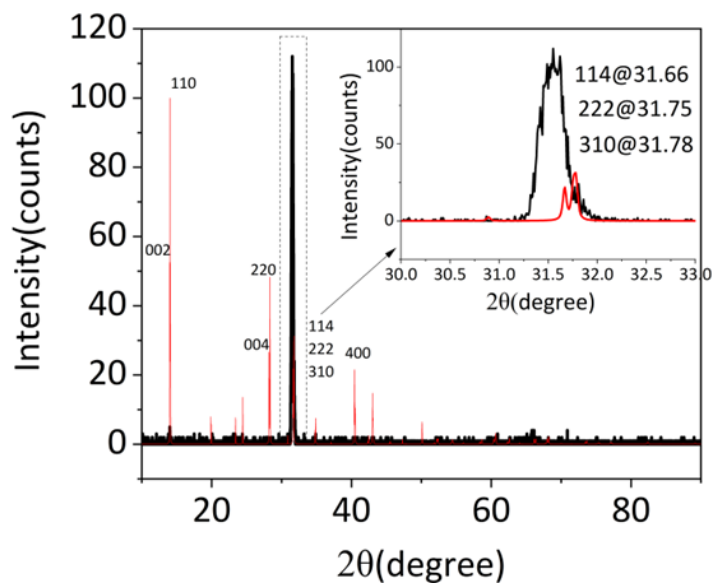

**Figure S10:** *XRD normal scan of n- type MAPI crystal along with superimposed powder simulation XRD simulation data, In the inset 31.6º peak is shown in zoom*

A $\theta - 2\theta$ scan of the n-MAPI single crystal shows a peak at 31.6°. This peak was compared with diffraction data extracted from the CIF file (card entry 7818931) for MAPI powder crystals. Figure S10 shows an overlay of the simulated data and the normal scan data of the MAPI single crystal, with an inset magnifying the 31.6° peak. High-resolution normal scans could not distinguish between the closely spaced three possible *hkl* planes (111, 114, 310).

A pole figure was chosen to identify the exact plane. The simulated pole figures for these planes, generated using WinWulff © is given in Figure S11. The experimentally observed pole pattern (shown in Figure 2(b) of the main text) closely aligns with the simulated pattern for the 310 plane in Figure S11(b). However, the center spot of the 310 pole-figure is tilted by 6.3° with respect to the substrate normal and exhibits a rotational φ shift of -43.6° relative to the substrate's longitudinal axis. Accounting for these angular offsets, the pole figure simulation was adjusted and plotted in Figure S12.

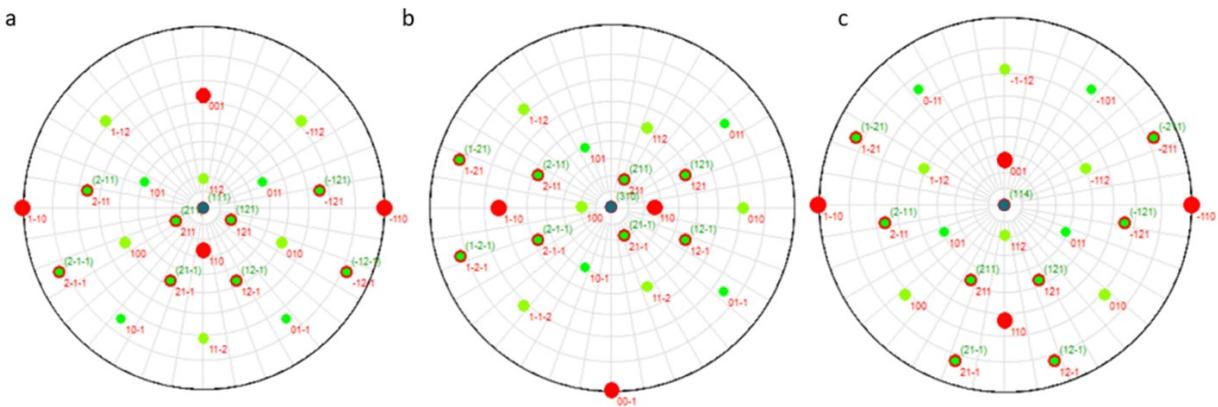

**Figure S11:** *Pole figure simulation of three probable planes (a) PF with respect to 111 plane (b) PF with respect to 310 plane (c) PF with respect to 114 plane is shown in encircled spots. Spotted <112> poles form a distinct pattern that differs.*

To validate the orientation, <110> plane poles were scanned and superimposed onto the <211> plane pole figure under the same XRD alignment conditions. This result precisely matched the simulated pattern in Figure S12. The plane poles exhibit lower intensity due to weaker 2θ peaks. Consequently, due to reduced interaction volume, the pole plot intensity diminishes significantly at higher χ (radial direction) in Figure 2(b).

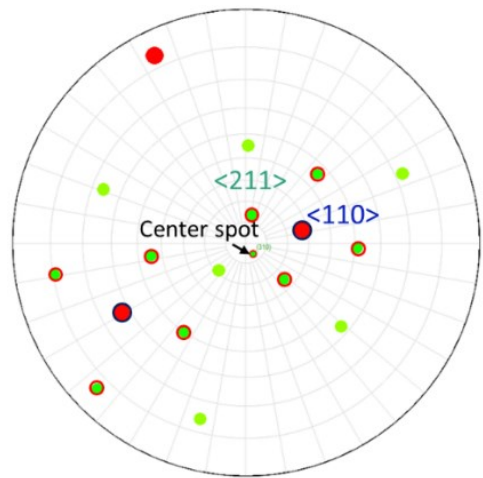

**Figure S12:** *Redrawn pole figure simulation with respect to 310 plane with exact center spot position considering χ offset and φ shift.*

# Steady-state and transient photoluminescence

Time-resolved photoluminescence (TRPL) measurements were performed using an FLS100 Edinburgh instrument. The MAPI single crystal sample was excited with a pulsed diode laser emitting at 450 nm, delivering a photon flux of $5 \times 10^{18}$ cm$^2$s$^{-1}$ (fluence ~10 µJ/cm$^2$/pulse) with a pulse period of 800 ns. A PMT 900 coupled with a monochromator was used for detection. The decay of the PL emission at 769 nm was collected for the MAPI single crystal using time-correlated single-photon counting (TCSPC) detection.

The same sample was excited with a continuous-wave 400 nm laser for steady-state measurements, providing an excitation photon flux of $8 \times 10^{17}$ cm$^{-2}$s$^{-1}$ (average power 10 mW). The PL emission spectra were scanned with a 2 nm slit width. The normalized PL spectra are shown in Figure 2d (main text). Analysis of the TRPL data revealed a bi-exponential decay behavior, following PL decay y = A1*exp(-x/τ1) + A2*exp(-x/τ2) + y$_0$ At high PL count rates, a bimolecular recombination lifetime of approximately 7 ns was observed. In the tail of the decay, where the signal enters the linear regime, a monomolecular recombination lifetime of 129 ns was determined (Figure S13).

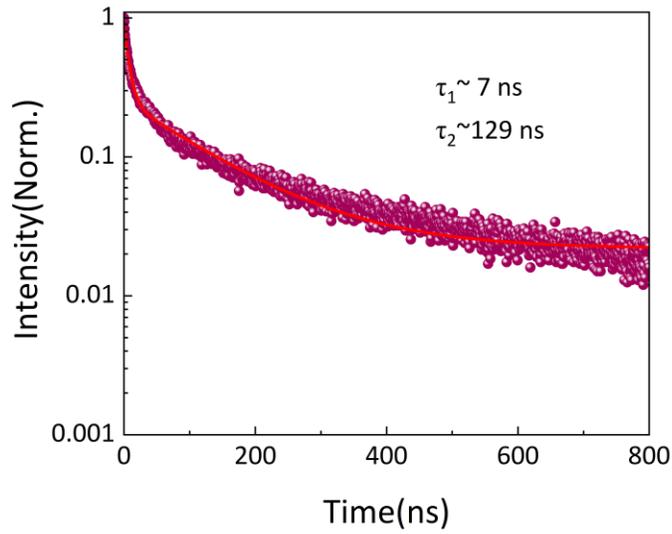

**Figure S13:** *The TR-PL of single crystal MAPI sample at carrier injection concentration $N_0 \sim 10^{16}$ cm$^{-3}$, in the inset normalized PL data is shown*

## MAPI Single crystal two probe I-V measurement

Before performing photo-AC Hall measurements on the MAPI single crystal, two- probe I-V characterization was conducted to assess the contact quality at all contacts. Figure S14 presents the current voltage sweep data for longitudinal and lateral contacts.

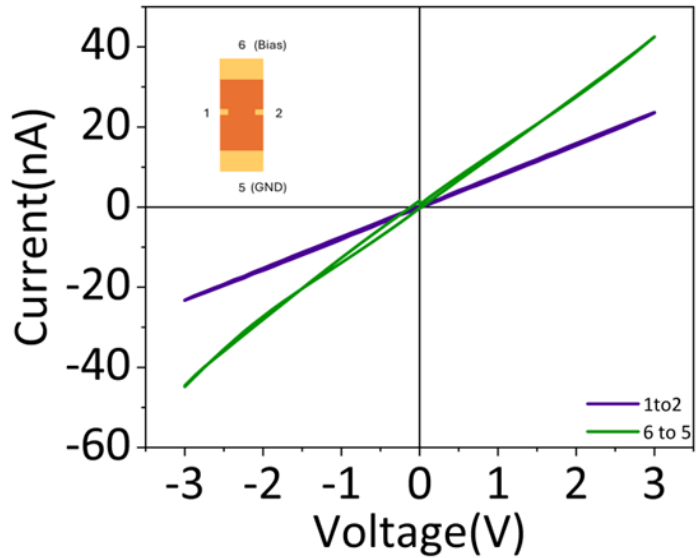

**Figure S14:** *Two probe Contact test for the MAPI crystal at dark condition*

Linearity in the current-voltage sweep across all contacts confirmed the ohmic nature of the contacts. Furthermore, the two-probe current was measured at various intensities during Hall measurements. As shown in Figure S15, the current exhibited a linear dependence on the applied bias ($V_{ap}$) across all measured intensities. This observation strongly suggests negligible contact resistance in the device.

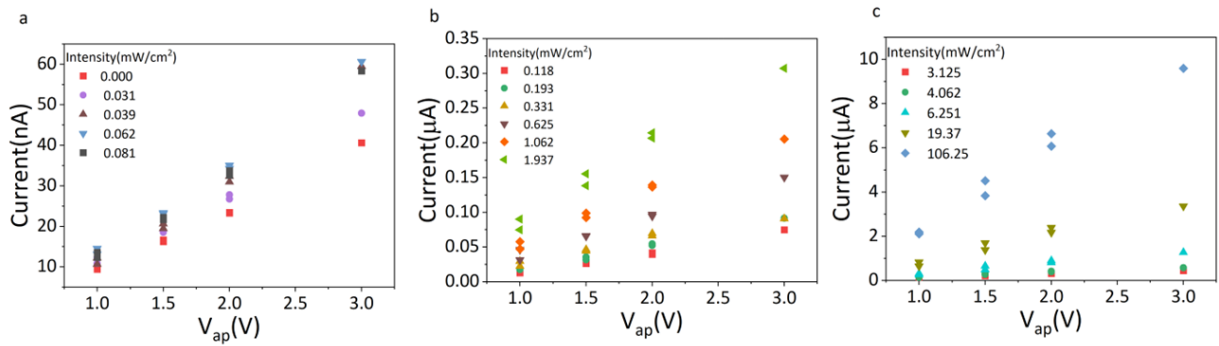

**Figure S15:** *Longitudinal Current as a function of applied bias at different intensities during Hall measurement.*

# Extracting carrier density using numerical iteration

In the main text, equation 3 fits the experimental $\mu_{PhotoHall}$ vs $\Delta p$ data obtained from photo hall and photoconductivity measurement. Here, $\mu_e$, $\mu_h$ and $n_o$ are the fitting parameters. The value obtained from the first iteration calculates the new $\Delta p^1$. Further, the same steps are followed to calculate the next iteration cycle value. This process continues until convergence.

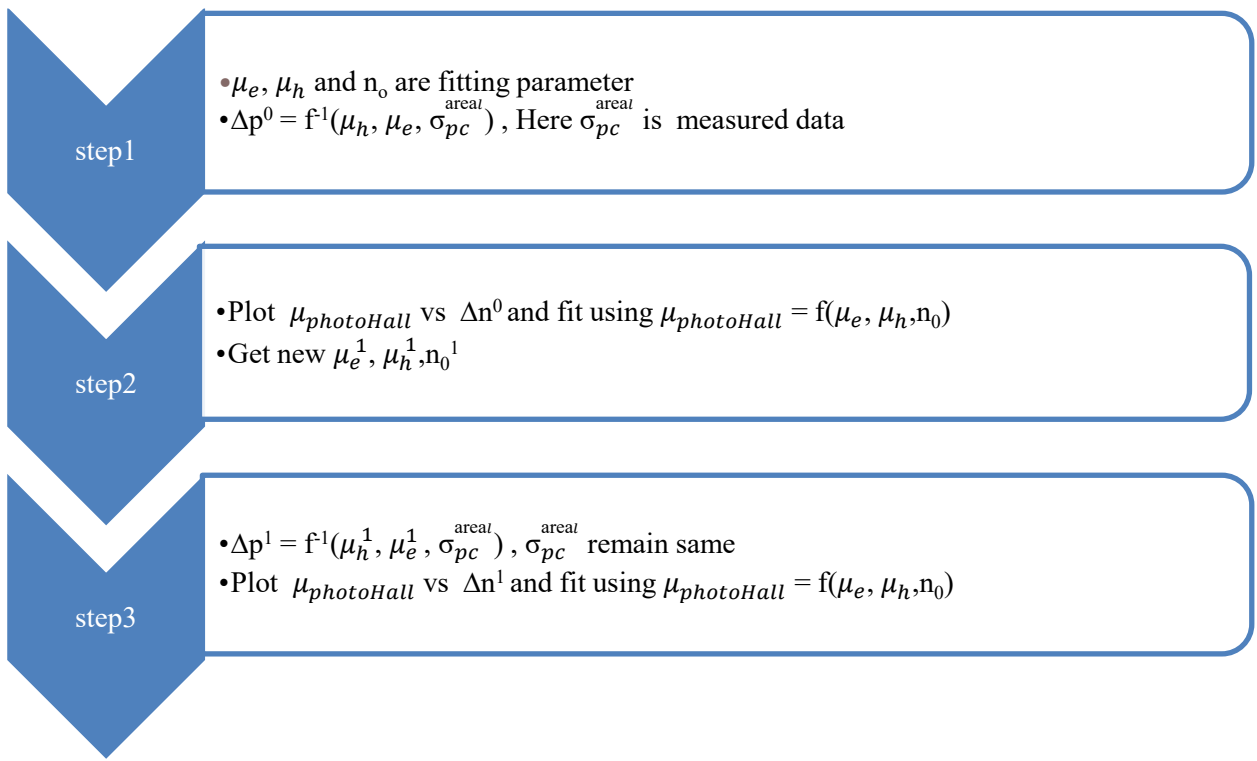

step1
- $\mu_e$, $\mu_h$ and $n_o$ are fitting parameter
- $\Delta p^0 = f^{-1}(\mu_h, \mu_e, \sigma_{pc}^{areal})$, Here $\sigma_{pc}^{areal}$ is measured data

step2
- Plot $\mu_{photoHall}$ vs $\Delta n^0$ and fit using $\mu_{photoHall} = f(\mu_e, \mu_h, n_0)$
- Get new $\mu_e^1$, $\mu_h^1$, $n_0^1$

step3
- $\Delta p^1 = f^{-1}(\mu_h^1, \mu_e^1, \sigma_{pc}^{areal})$, $\sigma_{pc}^{areal}$ remain same
- Plot $\mu_{photoHall}$ vs $\Delta n^1$ and fit using $\mu_{photoHall} = f(\mu_e, \mu_h, n_0)$

Figure S16 shows excess carrier concentration estimated from the final fit and data directly curated from the photoconductivity experiment.

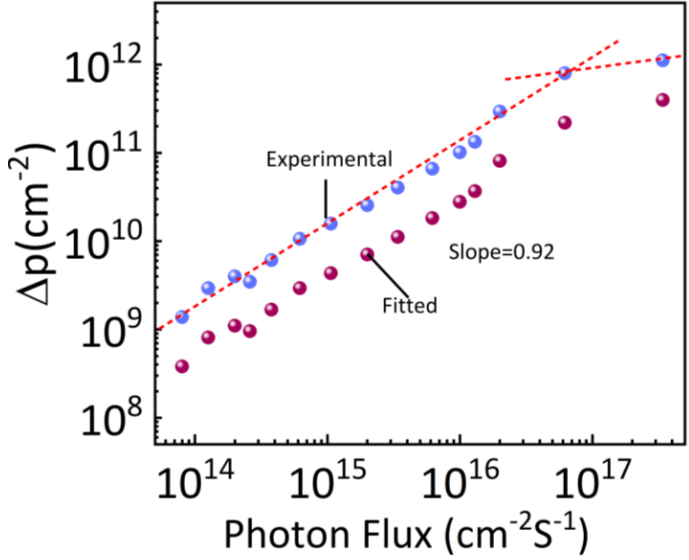

**Figure S16:** *Relation of excess carrier concentration with respect to photon flux*

## Error Estimation

### Error in $V_{Hall}$

$$V_{Hall} = \sqrt{(V^X - V_\circ^X)^2 + (V^Y - V_\circ^Y)^2}$$

$$\Delta V_{Hall} = \frac{1}{2}\frac{\pm \Delta X \pm \Delta Y}{V_{Hall}}$$

Where, $\pm\Delta X = 2(\Delta V^X + \Delta V_\circ^X)(V^X + V_\circ^X)$ and $\pm\Delta Y = 2(\Delta V^Y + \Delta V_\circ^Y)(V^Y + V_\circ^Y)$ are the standard deviations of the lock-in signals captured in the Hall measurement. Its standard deviation in X, and Y components in lock-in signal under bias and no bias are $\Delta V^X, \Delta V^Y, \Delta V_\circ^X, \Delta V_\circ^Y$, respectively. Here, errors are added in those places where they cancel each other, and higher-order errors are neglected.

### Error in Hall resistance

Hall resistance ($R_{xy}$) measures the slope of Hall voltage as a function of Hall current flows through the device under applied bias. Slopes are linear fit passing through the origin. The linear fit error is taken as an error for Hall resistance.